\documentclass[12pt]{article}

\usepackage{array} 
\usepackage{epsfig}
\usepackage{amssymb}
\usepackage{amsmath}                       
\usepackage{graphics,graphpap}
\usepackage{graphicx}
\usepackage{epstopdf}

\renewcommand{\thefootnote}{\fnsymbol{footnote}}

\begin{document}

\begin{titlepage}

\vskip1.5cm
\begin{center}

{\Large \bf \boldmath sFit:  a method for background subtraction in maximum likelihood fit}

    \vskip1.3cm {
     Yuehong Xie\footnote{Yuehong.Xie@cern.ch}
\vskip0.3cm    
 {\em  University of Edinburgh, Edinburgh EH9 3JZ, United Kingdom }}
    \vskip0.5cm 


\vskip5cm

{\large\bf Abstract\\[10pt]} \parbox[t]{\textwidth}{

This paper presents a statistical method to subtract background in maximum likelihood fit, without 
relying on any separate sideband or simulation for background modeling. 
The method, called sFit, is an extension to the sPlot technique 
originally  developed to reconstruct true distribution for each date component.
The sWeights defined for the sPlot technique allow to construct a modified likelihood function using only the signal 
probability density function and events in the signal region. Contribution of background events in
the signal region to the likelihood function cancels out on a statistical basis.
Maximizing this likelihood function leads to unbiased estimates of the fit parameters in the signal probability density function.

}

\vfill

\end{center}
\end{titlepage}

\setcounter{footnote}{0}
\renewcommand{\thefootnote}{\arabic{footnote}}

\newpage

\section{Introduction}
The method of maximum likelihood  is a common procedure used for parameter estimation in analysis of experimental data.
Suppose the probability density function (pdf) $P(x;\theta)$ with unknown parameters $\theta$ describes a set of $N$
independent measurements $x_e$. The values of $\theta$ that maximize the likelihood function 
\begin{equation}
\label{eq:likelihood}
 L(\theta) =   \prod_{e=1}^N P(x_e;\theta)
\end{equation}
are taken to be the estimators for  $\theta$. 
The conventional method to include background in maximum likelihood fit requires to  write the total pdf as 
\begin{equation}
P (x;\theta,f_s) = f_s P_s (x;\theta) + (1-f_s) P_b (x)
\end{equation}
where $f_s$ is the fraction of signal events in th data sample, $P_s(x;\theta)$ and $P_b (x)$ are the signal and 
background pdf respectively. 
Usually the  background pdf  $P_b (x)$  needs to be  obtained from either  Monte Carlo simulation or
separate sidebands. The latter requires to divide data into signal region and sidebands using 
discriminating variables $y$ which are supposed to be uncorrelated with $x$ for the background component.
Some problems may arise with this method:  $P_b (x)$ may be too complicated to parameterize; 
the parameterization of $P_b (x)$ obtained from simulation may be unreliable;
the sidebands may have very different $P_b (x)$ distributions from the signal region  
if they  are too far away from the signal region;
the sidebands may contain a significant signal component if they are too close to the signal region.
Therefore, it is highly desirable to have an alternative  method which does not rely on 
background parameterization from either simulation or separate sidebands. This paper provides a solution
by generalizing the sPlot technique~\cite{splot}, originally  developed to reconstruct true distribution of $x$ for 
the signal component using sWeights defined as functions of $y$, into a modified maximum likelihood method, called sFit.

\section{ The sFit method}
\label{sec:method}

Suppose $x$ are uncorrelated with the  discriminating variables $y$, i.e. the distribution of $x$ is independent of 
$y$, for both signal and   background components\footnote{This condition is easier to satisfy for a smaller signal region in $y$.}.
The data sample contains $N_s$ signal events and $N_b$ background events. The distributions of $y$ for signal and background
are denoted as  $F_s(y)$ and $F_b(y)$ respectively. We assume that $N_s$, $N_b$, $F_s(y)$ and $F_b(y)$ are known.
Following the  formalism of the sPlot technique,  we define a sWeight function for the signal component:
\begin{equation}
\label{eq:ws}
W_s(y) = { V_{ss} F_s(y) +V_{sb} F_b(y)   \over N_s F_s(y) + N_b F_b(y)},
\end{equation}
where the matrix $V$ is obtained by inverting the matrix 
\begin{equation}
V_{ij}^{-1} = \sum_{e=1}^{N} { F_i(y_e)F_j(y_e) \over (N_s F_s(y_e)+N_b F_b(y_e) )^2}.  
\end{equation}
The sWeight for each event, $W_s(y_e)$, can be calculated.  The basic idea of the sPlot technique is that 
the histogram of $x_e$ weighted by $W_s(y_e)$ represents the true distribution of $x$ for the signal component,
because the background contribution to the histogram cancels out due to this choice of $W_s(y_e)$.

We can extend this idea one step further and define a weighted likelihood function
\begin{equation}
 L_W(\theta) =   \prod_{e=1}^N [P_s(x_e;\theta)]^{W_s(y_e)}.
\end{equation}
We expect that background contribution to this likelihood function $L_W(\theta)$  cancels on a statistical basis,
therefore $\theta$ can be estimated by maximizing $L_W(\theta)$.

\section{ Application}

We apply the sFit method in a simple case in time-dependent analysis of B decays.
The signal pdf of proper time $t$ is
\begin{equation}
P_s(t;A, \Gamma) = C_1 e^{-\Gamma t}(1+A\sin(\Delta m t))
\end{equation}
where $C_1$ is a normalization factor, $\Delta m = 17$ ${\rm ps}^{-1}$ is known, $A$ and $\Gamma$ are 
parameters to be determined. Dependence of the pdf on initial flavour of the B meson  is not considered 
for simplicity. The background events have a different time distribution
\begin{equation}
P_b(t; \Gamma_b) = C_2 e^{-\Gamma_b t},
\end{equation}
where $C_2$ is a normalization factor, and $\Gamma_b$ is unknown.
 The discriminating variable is the B mass. The signal events have a known gaussian 
mass distribution with a standard deviation $\sigma_m = 15$ $ {\rm MeV}$ and mean $m_0 = 5369$ ${\rm MeV}$.
The background events have a known flat distribution. The signal mass window is chosen to be centered
at $m_0$. We consider scenarios with different half mass window size $S_m$ and different number of signal and 
background events in the signal mass window.

\begin{figure}[ht]
\vfill\begin{minipage}{0.5\linewidth}
   \includegraphics[angle=90,width=80mm]{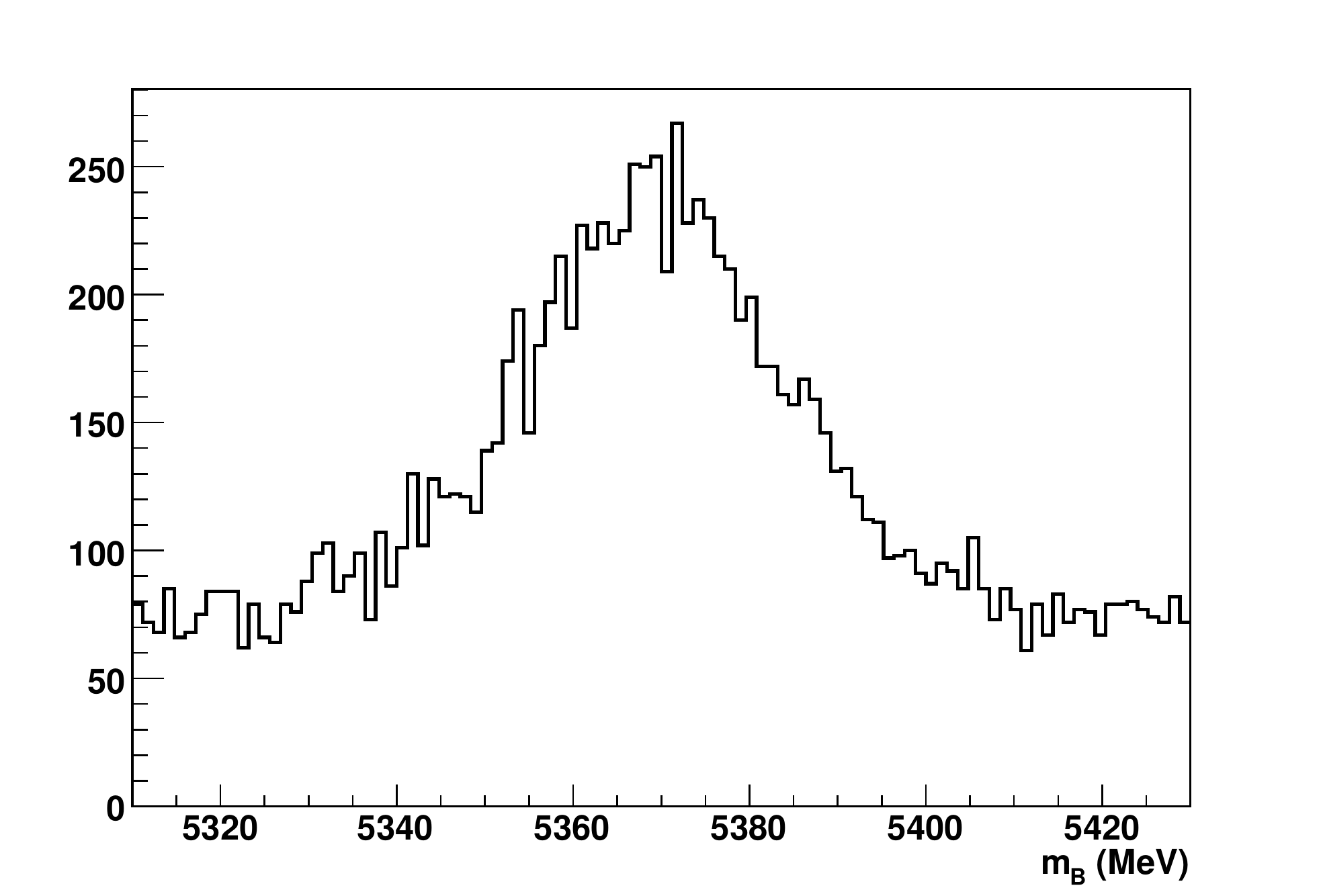}
\end{minipage}
\begin{minipage}{0.5\linewidth}
   \includegraphics[angle=90,width=80mm]{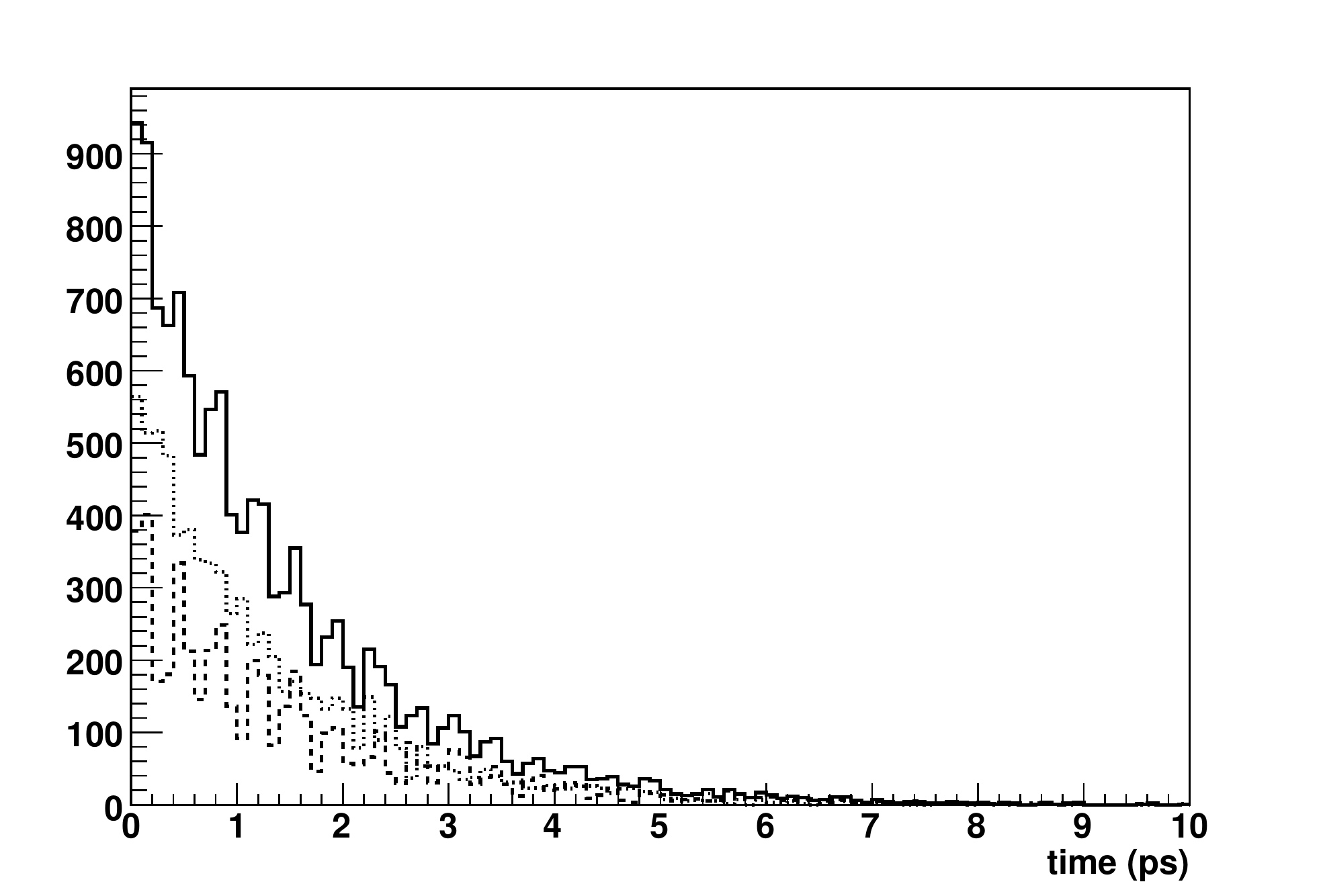}
\end{minipage}
\caption{Distributions from a data set with  $S_m/\sigma_m=6$, $N_s= 5000$ and $N_b/N_s =1.5$.
Left: the B mass distribution; right: the total time distribution (solid), as well as the signal time distribution (dashed) 
and background time distribution (dot-dashed) reconstructed using the sPlot technique.   
}
\label{fig:mtdist}
\end{figure}

\begin{figure}
\centering   \includegraphics[angle=90,width=80mm]{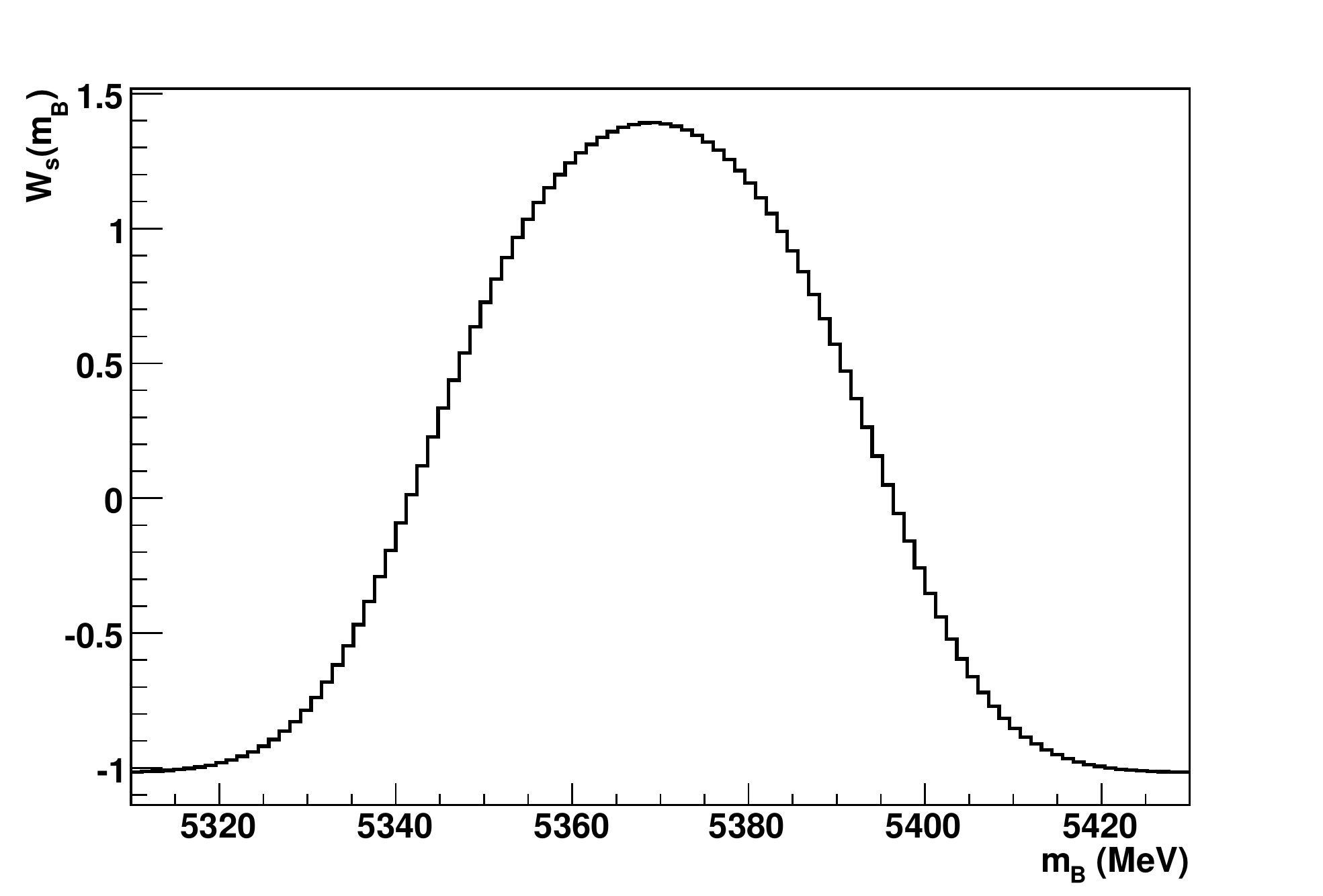}
\caption{$W_s$ as a function of the B mass $m_B$ for the scenario  $S_m/\sigma_m=6$, $N_s= 5000$ and $N_b/N_s =1.5$. }
\label{fig:ws}
\end{figure}

As an example,  Figure~\ref{fig:mtdist} shows the distribution of the B mass $m_B$, the total time distribution,
as well as the signal and background time distributions reconstructed using the sWeights,
for the scenario  $S_m/\sigma_m=6$, $N_s= 5000$ and $N_b/N_s =1.5$.
The $W_s(m_B)$ function for this scenario is shown in Figure~\ref{fig:ws}.
The fact that $W_s(m_B)$ has positive values in the high signal purity region around $m_0$ and negative values 
in the low purity area illustrates why the background contribution cancels in both the sPlot and the sFit methods.

500 toy data sets are generated for each scenario with 
\begin{equation}
 A= 0.5,\, \Gamma = 0.65\, {\rm ps}^{-1},\, \Gamma_b = 0.8\, {\rm ps}^{-1}.
\end{equation}
 We perform fit to each data set
using two different methods: the sFit method described in Section~\ref{sec:method} and a conventional  
maximum likelihood method for reference based on Equation~\ref{eq:likelihood}  and the total pdf 
\begin{equation}
 P (t;A, \Gamma, \Gamma_b) = f_s P_s (t;A, \Gamma) F_s(m) + (1-f_s) P_b (t; \Gamma_b)  F_b(m)
\end{equation}
where the shape of the background pdf $P_b(t;\Gamma_b)$ is assumed to be known except the parameter $\Gamma_b$.

\begin{figure}[ht]
\vfill\begin{minipage}{0.5\linewidth}
   \includegraphics[angle=90,width=80mm]{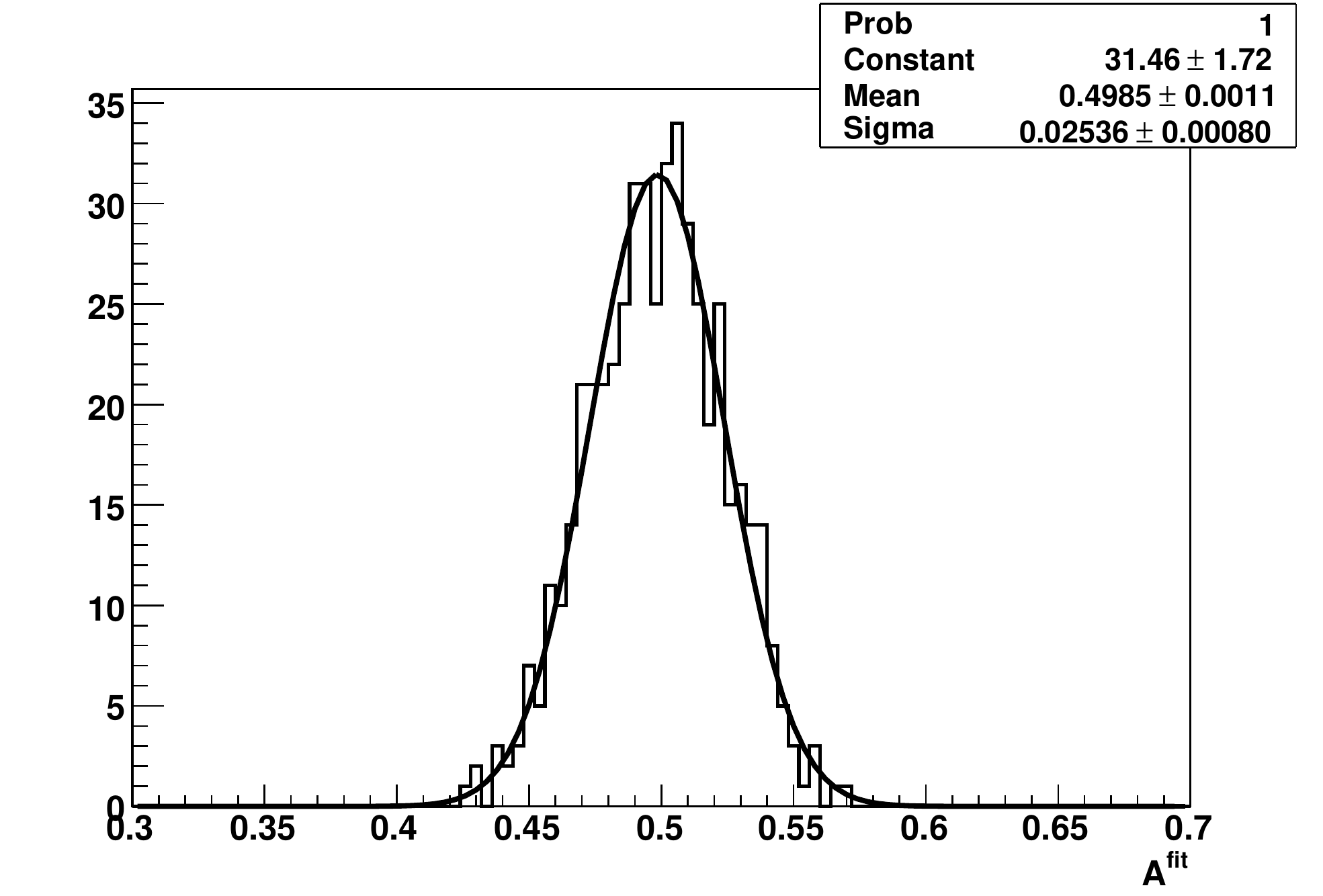}
\end{minipage}
\begin{minipage}{0.5\linewidth}
   \includegraphics[angle=90,width=80mm]{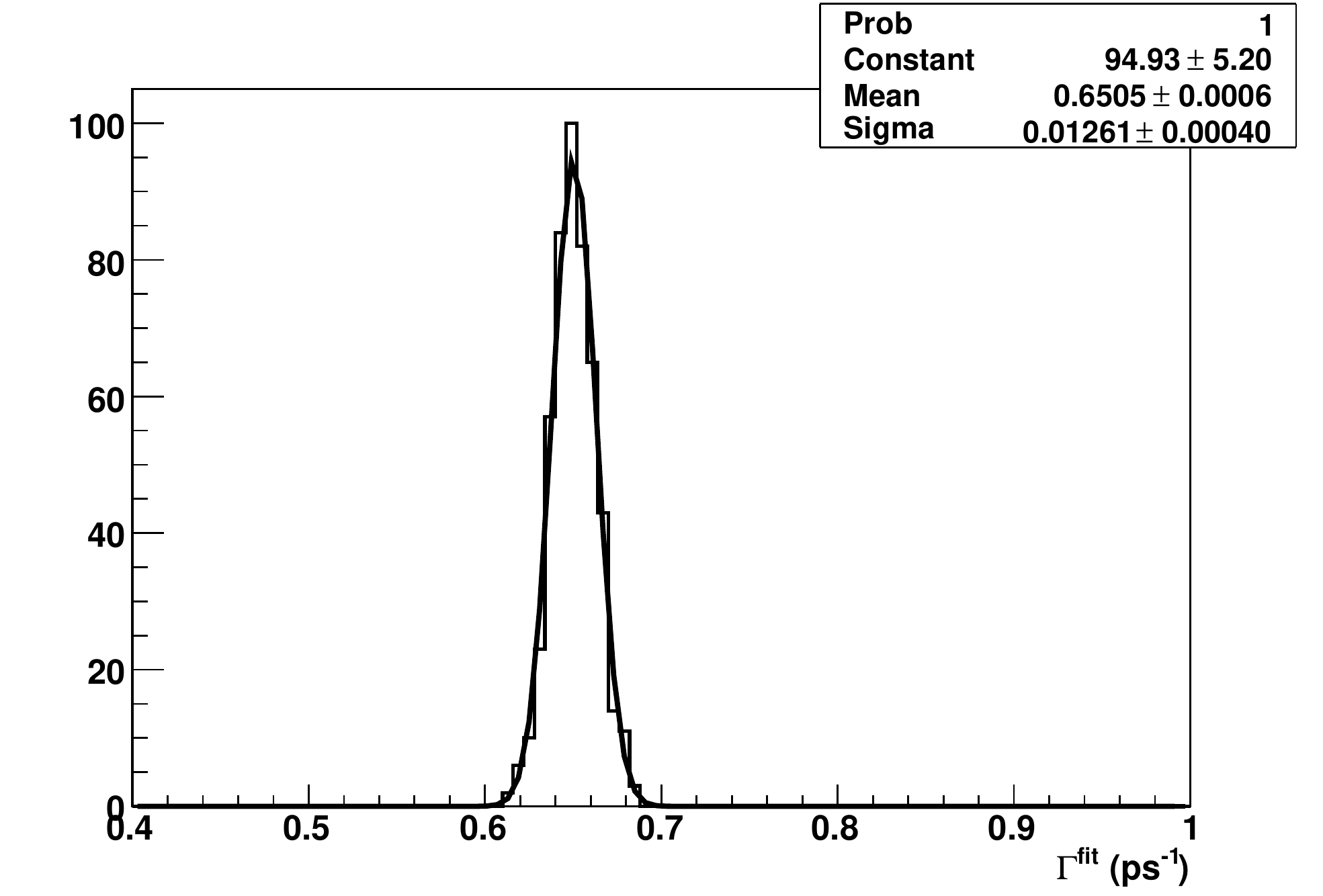}
\end{minipage}
\caption{Distributions of the estimated values of $A$ (left) and $\Gamma$ (right)
obtained with the sFit method, with superimposed gaussian fits, 
 for the scenario $S_m/\sigma_m=6$, $N_s= 5000$ and $N_b/N_s =1.5$. 
 The input values are $A= 0.5$ and $\Gamma = 0.65\,{\rm ps^{-1}}  $.
}
\label{fig:sfit}
\end{figure}

\begin{figure}[ht]
\vfill\begin{minipage}{0.5\linewidth}
   \includegraphics[angle=90,width=80mm]{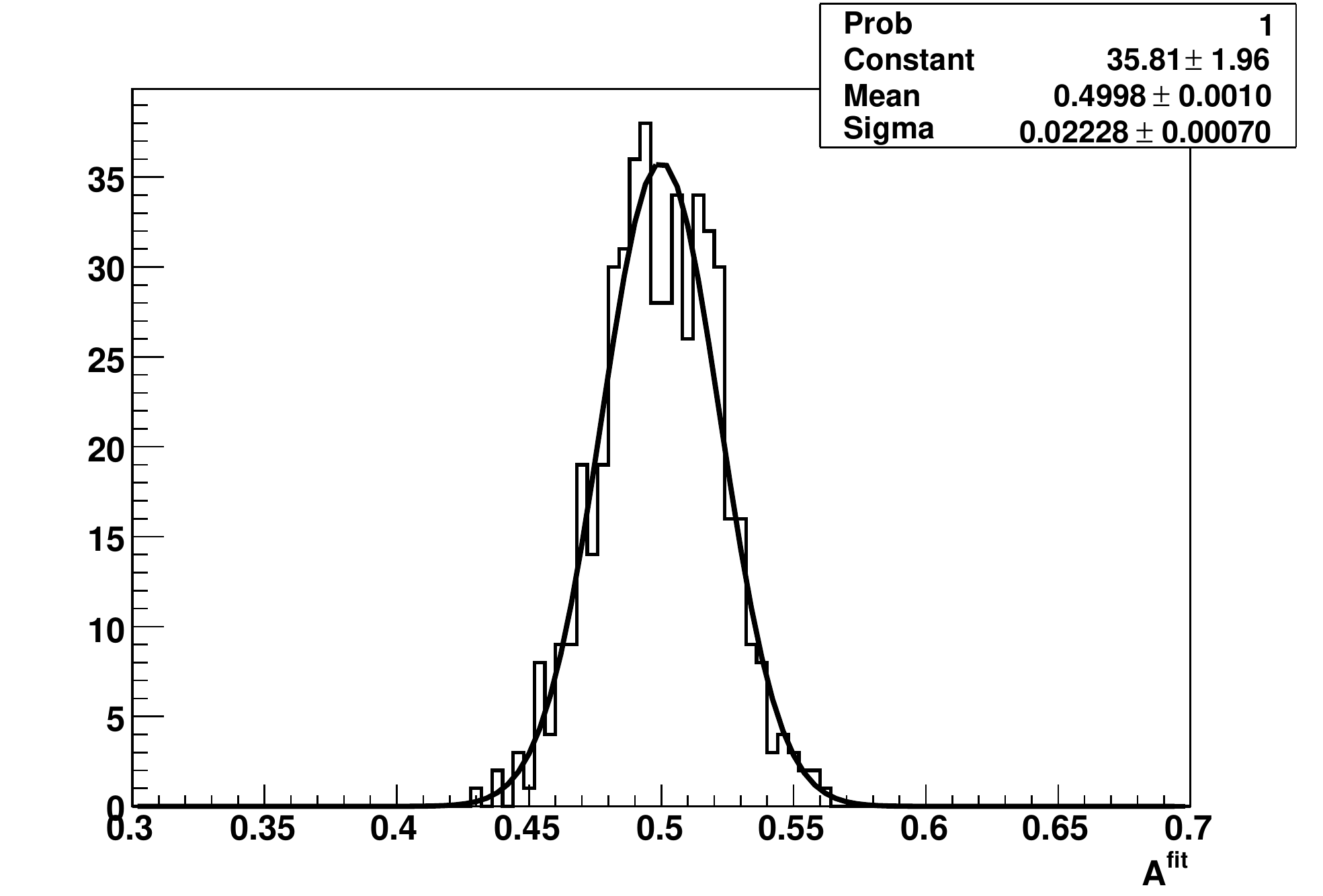}
\end{minipage}
\begin{minipage}{0.5\linewidth}
   \includegraphics[angle=90,width=80mm]{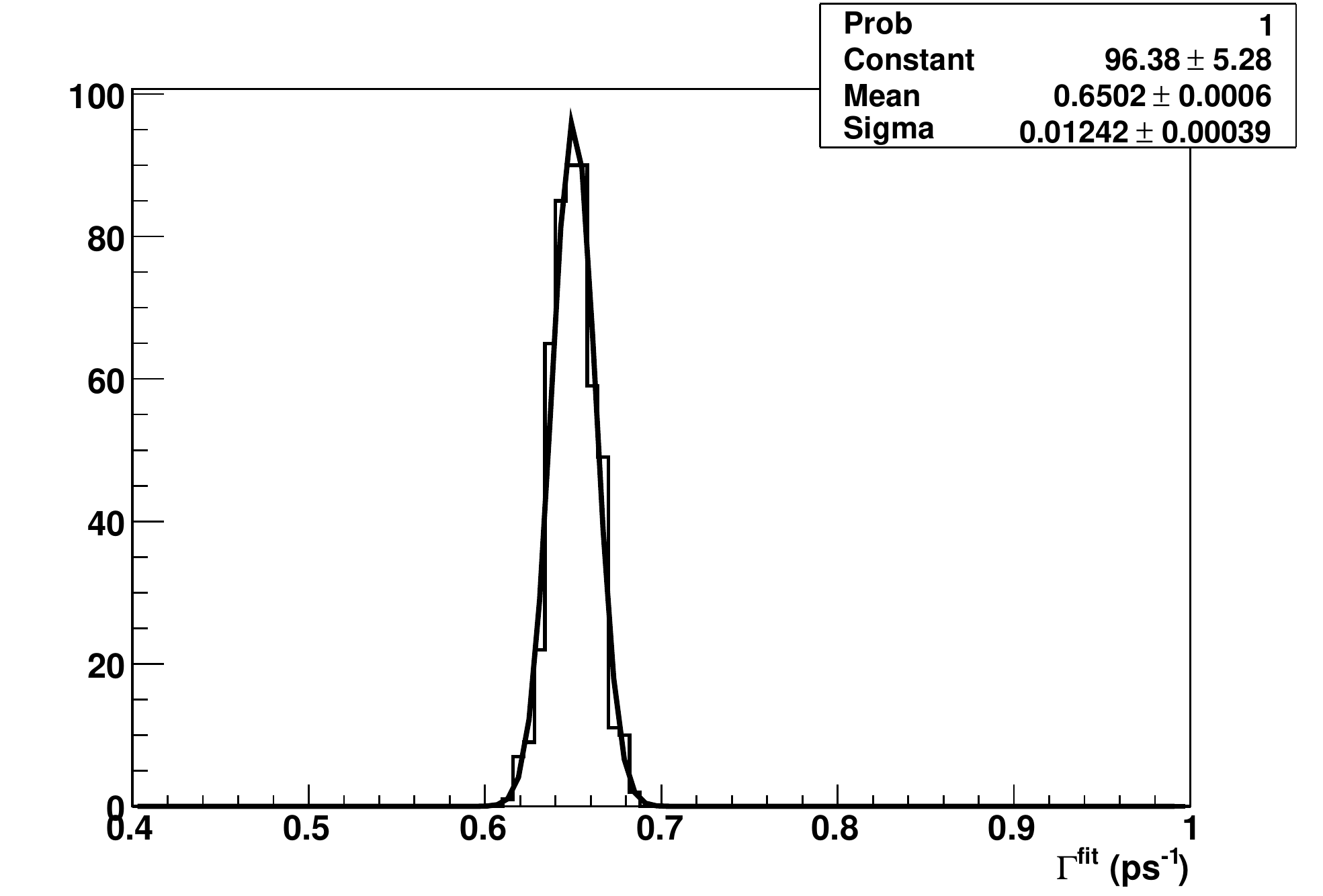}
\end{minipage}
\caption{Distributions of the estimated values of $A$ (left) and $\Gamma$ (right)
obtained with the reference method, with superimposed gaussian fits,
 for the scenario $S_m/\sigma_m=6$, $N_s= 5000$ and $N_b/N_s =1.5$. 
 The input values are $A= 0.5$ and $\Gamma = 0.65\,{\rm ps^{-1}}  $.
}
\label{fig:mlfit}
\end{figure}

For each scenario and each fit method, the statistical errors and mean values of the parameter $A$ and $\Gamma$ 
are obtained using a single gaussian fit to their estimated values from  the 500 data sets.
An example is shown in Figure~\ref{fig:sfit} and Figure~\ref{fig:mlfit}
for the scenario $S_m/\sigma_m=6$, $N_s= 5000$ and $N_b/N_s =1.5$. 
The results for different  scenarios  are summarized in the Table~\ref{tab:res1} and Table~\ref{tab:res2}
for the sFit method and reference method  respectively. 

It can be seen that the sFit method gives unbiased estimates of the fit parameters $A$ and $\Gamma$. The statistical
errors of the parameter  estimates obtained with  the sFit  method are bigger than 
the errors of the estimates obtained with the reference methods. This has two reasons: while the background contribution
 to $L_W$ cancels, part of the signal contribution is also lost due to the negative sWeights  in the low purity area, 
and the size of the loss depends on the size of the signal mass  window;
the cancellation of background contribution  is not exact due to statistical fluctuation, and the size of the fluctuation depends
on the background level in the signal region. 
In general, the larger the signal mass window, the smaller the precision difference  between the two methods; 
the lower the background level,  the smaller the precision difference  between the two methods. We should keep in mind 
that the reference method takes full advantage of the knowledge of the  background time distribution, which is usually 
unavailable  or unreliable in real data analysis, therefore the parameter errors in Table~\ref{tab:res2}  are 
too optimistic and should be
regarded as lower  limits rather than realistic estimates.
The sWeight function defined in Equation~\ref{eq:ws} is not the unique way to define event weight function in order 
to can cancel
the background contribution to the weighted likelihood function. It would be interesting to investigate if the sWeight 
is the optimal choice of event weight function that minimizes the parameter errors.

\begin{figure}[ht]
\vfill\begin{minipage}{0.5\linewidth}
   \includegraphics[angle=90,width=80mm]{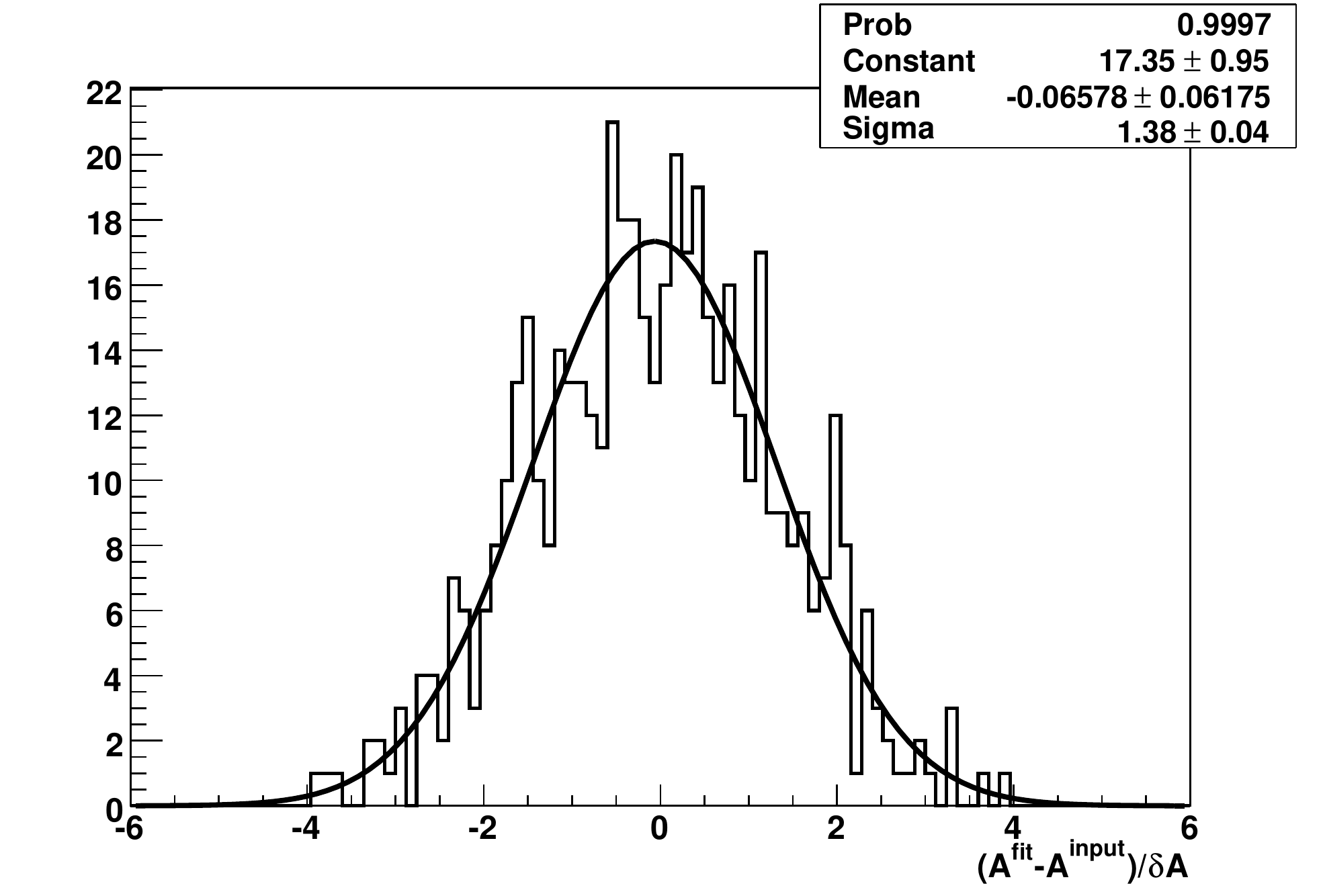}
\end{minipage}
\begin{minipage}{0.5\linewidth}
   \includegraphics[angle=90,width=80mm]{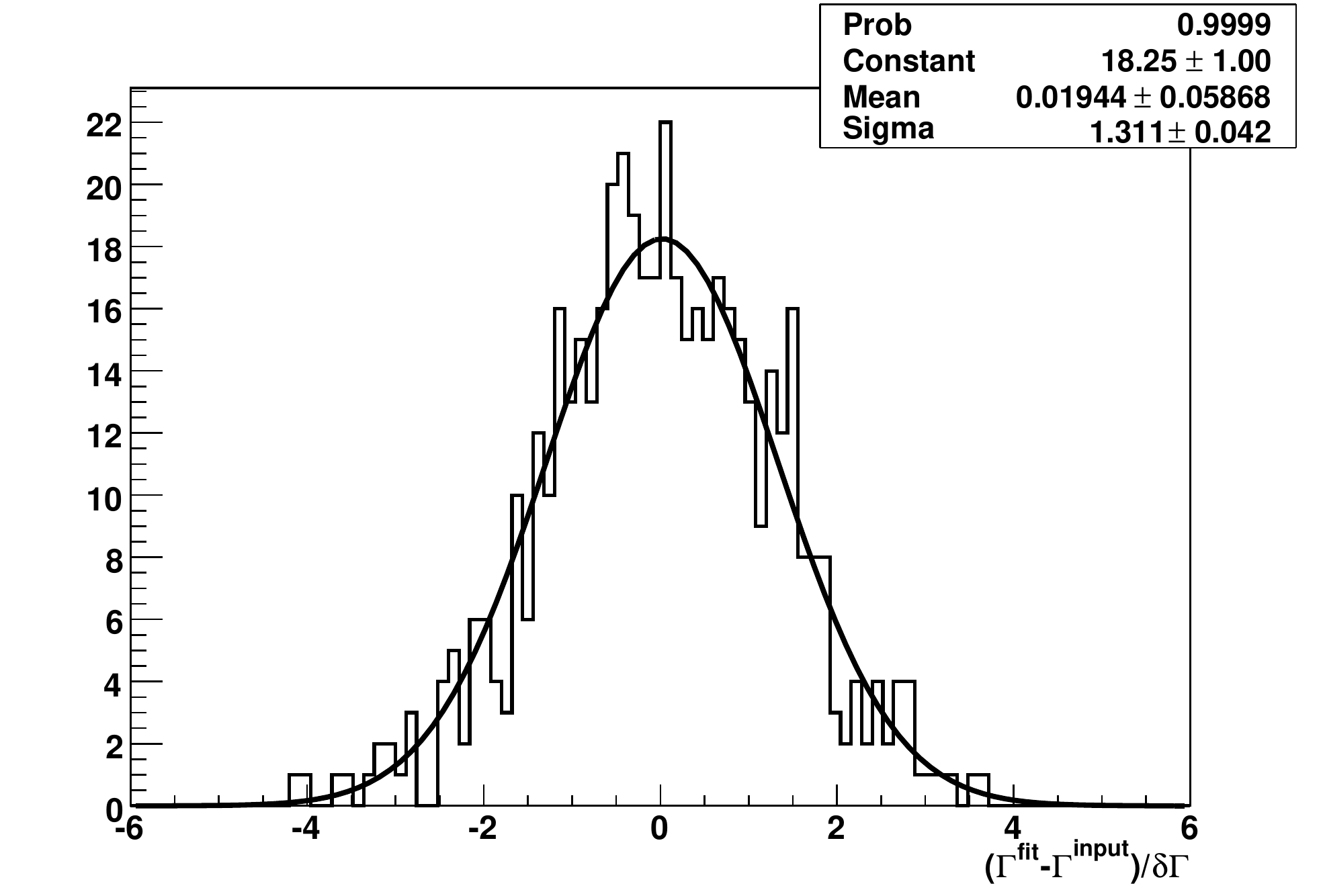}
\end{minipage}
\caption{Distributions of  $(A^{fit}-A^{input})/\delta A$ (left)
and $(\Gamma^{fit}-\Gamma ^{input})/\delta \Gamma$ (right)
obtained with the sFit method, with superimposed gaussian fits,
for the scenario $S_m/\sigma_m=6$, $N_s= 5000$ and $N_b/N_s =1.5$.
}
\label{fig:pull_sfit}
\end{figure}

\begin{figure}[ht]
\vfill\begin{minipage}{0.5\linewidth}
   \includegraphics[angle=90,width=80mm]{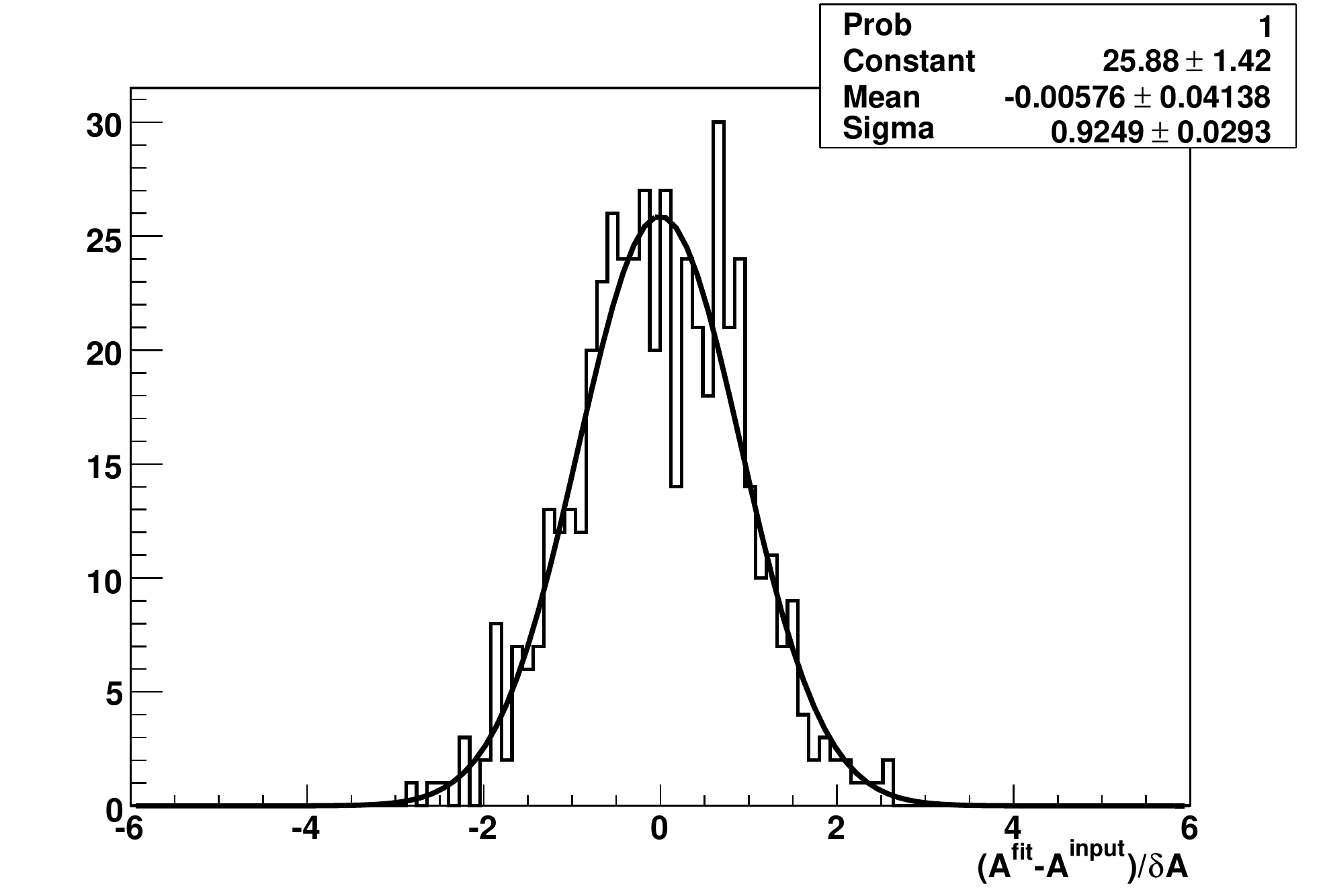}
\end{minipage}
\begin{minipage}{0.5\linewidth}
   \includegraphics[angle=90,width=80mm]{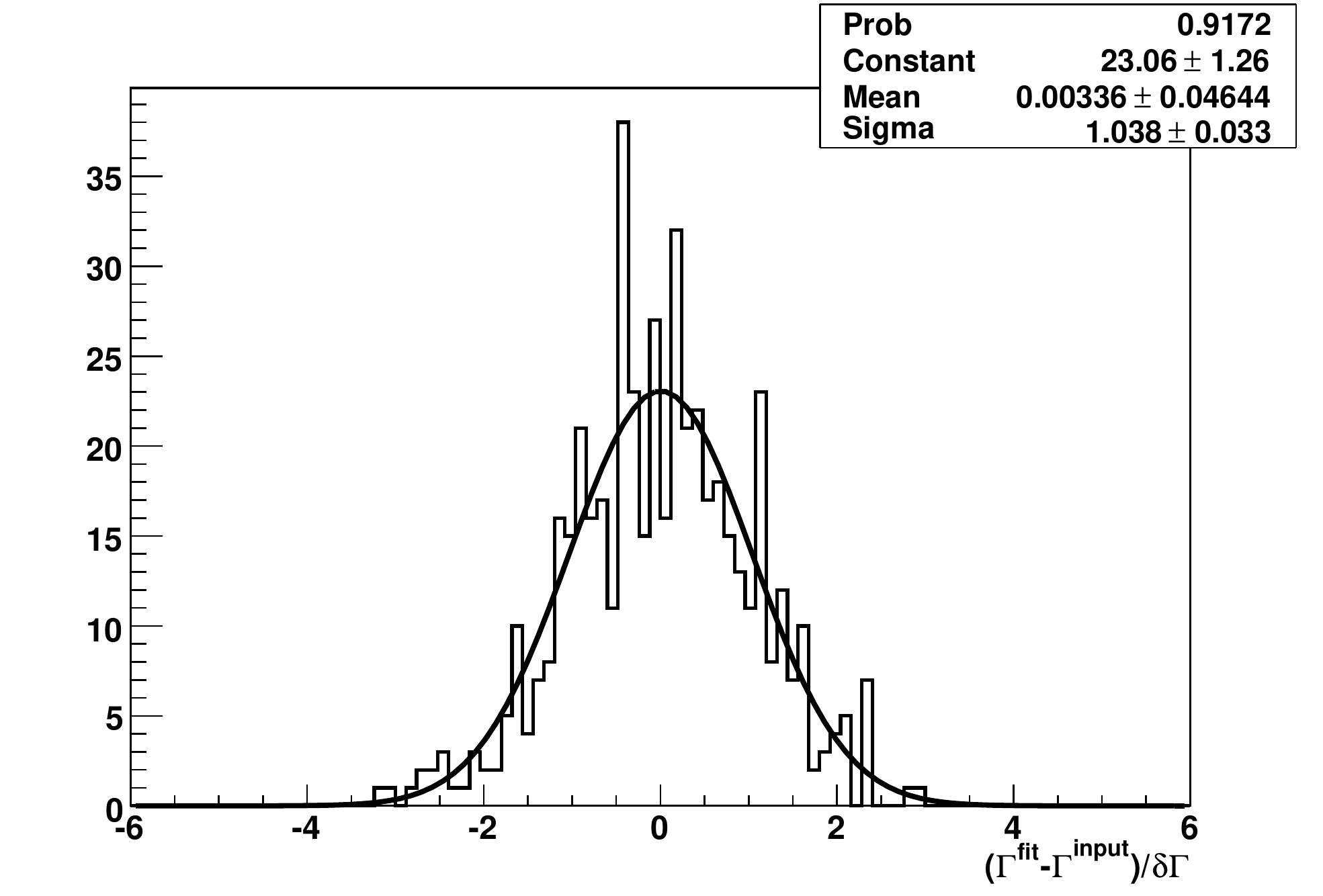}
\end{minipage}
\caption{Distributions of  $(A^{fit}-A^{input})/\delta A$ (left)
and $(\Gamma^{fit}-\Gamma ^{input})/\delta \Gamma$ (right)
obtained with the sFit method, with superimposed gaussian fits,
for the scenario $S_m/\sigma_m=6$, $N_s= 5000$ and $N_b/N_s =1.5$.
}
\label{fig:pull_mlfit}
\end{figure}

 Figure~\ref{fig:pull_sfit} and  Figure~\ref{fig:pull_mlfit}
 show  the distributions of $(A^{fit}-A^{input})/\delta A$ and $(\Gamma^{fit}-\Gamma ^{input})/\delta \Gamma$
 for the sFit method and reference method respectively, 
where $\delta A$ and $\delta \Gamma$ are the parameter errors estimated by the Minuit program 
according to $\ln L = \ln L_{max} - {1 \over 2}$. Apparently the errors obtained this way using the weighted
likelihood function $L_W$ are underestimated, because the effect of background fluctuation is not properly accounted 
for. Reliable error estimates can be obtained from Monte Carlo simulation.

\begin{table}[h]
\begin{center}
  \begin{tabular}{|c|c|c|c|c|}
 \hline
$S_m/\sigma_m$, $N_s$, $N_b/N_s$ & $\sigma(A)$ & mean of $A$  & $\sigma(\Gamma)$ (${\rm ps}^{-1}$)  & mean of $\Gamma$ (${\rm ps}^{-1}$)   \tabularnewline
 \hline
4, 5000, 1   & 0.0304  &  0.502   & 0.0134 & 0.6504 \tabularnewline
 \hline
6, 5000, 1.5   & 0.0254  &  0.498   & 0.0126 & 0.6504 \tabularnewline
 \hline
4, 5000, 0.5   & 0.0243   &  0.501   & 0.0115 & 0.6511 \tabularnewline
 \hline
6, 5000, 0.75   & 0.0223   &  0.501   & 0.0107 & 0.6496 \tabularnewline
 \hline

  \end{tabular}
\end{center}
\normalsize
\caption{Statistical errors and mean values of $A$ and $\Gamma$ from 500 fits using the sFit method
for different scenarios. Errors of the numbers are on the last digits.
 The input values are $A= 0.5$ and $\Gamma = 0.65\,{\rm ps^{-1}}  $.
}
\label{tab:res1}
\end{table}

\begin{table}[h]
\begin{center}
  \begin{tabular}{|c|c|c|c|c|}
 \hline
$S_m/\sigma_m$, $N_s$, $N_b/N_s$ & $\sigma(A)$ & mean of $A$  & $\sigma(\Gamma)$ (${\rm ps}^{-1}$)  & mean of $\Gamma$ (${\rm ps}^{-1}$)  \tabularnewline
 \hline
4, 5000, 1   & 0.0251  &  0.502   & 0.0129 & 0.6506 \tabularnewline
 \hline
6, 5000, 1.5   & 0.0223  &  0.500   & 0.0124 & 0.6502 \tabularnewline
 \hline
4, 5000, 0.5   & 0.0215   &  0.500   & 0.0113 & 0.6511 \tabularnewline
 \hline
6, 5000, 0.75   & 0.0211   &  0.501   & 0.0105 & 0.6496 \tabularnewline
 \hline

  \end{tabular}
\end{center}
\normalsize
\caption{Statistical errors and mean values of $A$ and $\Gamma$ from 500 fits using the conventional maximum likelihood method
for different scenarios. Errors of the numbers are on the last digits.
 The input values are $A= 0.5$ and $\Gamma = 0.65\,{\rm ps^{-1}}  $.
}
\label{tab:res2}
\end{table}

\clearpage

\section{Conclusions}

The sFit method presented in this paper fully exploits the idea of background cancellation
for maximum likelihood fit. If the variables $x$ are uncorrelated with the
discriminating variables $y$ for both signal and background components,
one can define an event  weight function of $y$ which can  be used 
not only to reconstruct the signal distribution of $x$, 
but also for parameter estimation from the distributions of $x$ 
in  maximum  likelihood fit without explicitly modeling the background.
 The likelihood function constructed 
using the event weights and the signal pdf of $x$ is free
from background contribution on a statistical basis.
Maximizing the likelihood function leads to unbiased parameter estimates.
This method can largely reduce systematic uncertainties due to unreliable
background model obtained from either sidebands or simulation at a cost of
modest increases in the statistical errors.



\end{document}